\documentclass[%
 amsmath,amssymb,
 aps,notitlepage
]{revtex4-1}

\usepackage{graphicx}
\usepackage{bm}

\usepackage{amsthm,amsfonts,amssymb,epsfig,graphics,amsmath,oldgerm,color}

\usepackage[latin1]{inputenc} 

\newcommand{\R}{{\mathbb R}}

\newcommand{\Z}{{\mathbb Z}}

\def\veps{\varepsilon}

\newcommand{\eps}{\varepsilon}
\newcommand{\OO}{\mathcal{O}}
\newcommand{\pa}{\partial}
\newcommand{\bspm}{\left(\begin{smallmatrix}}\newcommand{\espm}{\end{smallmatrix}\right)}
\newcommand{\bpm}{\left(\begin{matrix}}\newcommand{\epm}{\end{matrix}\right)}

\newcommand{\PROOF}{\textbf{Proof.} }

\def\epsilon{\varepsilon}
    
\def\beq{\begin{equation}}
\def\eeq{\end{equation}}

\begin{document}


\title{Phononic Rogue Waves}


\author{E.~G. Charalampidis, J.~Lee and P.~G.~Kevrekidis}
\affiliation{Department of Mathematics and Statistics, University of Massachusetts \\
Amherst MA 01003-9305, USA}%

\author{C. Chong}
\email{cchong@bowdoin.edu}
 \affiliation{Department of Mathematics, Bowdoin College, Brunswick, ME 04011, USA}

\date{\today}

\begin{abstract}
We present a theoretical study of extreme events occurring in 
phononic lattices. In particular, we focus on the formation of
rogue or freak waves, which are characterized by their localization
in both spatial and temporal domains. We consider two examples. 
The first one is the prototypical nonlinear mass-spring system in the
form of a homogeneous Fermi-Pasta-Ulam-Tsingou (FPUT) lattice with
a polynomial potential.  By deriving an approximation based on the
nonlinear Schr\"odinger (NLS) equation, we are able to initialize 
the FPUT model using a suitably transformed
Peregrine soliton solution of the NLS,
obtaining dynamics that resembles a rogue wave on the FPUT lattice. We also 
show that Gaussian initial data can lead to dynamics
featuring rogue wave for
sufficiently wide Gaussians. The second example is a diatomic granular 
crystal exhibiting rogue wave like dynamics, which we also obtain through
an NLS reduction and numerical simulations. The granular crystal (a 
chain of particles that interact elastically) is a widely studied 
system that lends itself to experimental studies. This study 
serves to illustrate the potential of
such dynamical lattices
towards the experimental observation of acoustic rogue waves.
\end{abstract}



\pacs{Valid PACS appear here}
\maketitle




\section{Introduction}

Extreme wave events, such as freak or rogue waves, are waves 
that seem to appear out of nowhere, and then vanish without a
trace \cite{r1,r2,r3}. The term rogue wave was first coined 
to describe an ocean wave that has an amplitude greater than 
twice the significant wave height \cite{r1}. Based on the classical 
description of waves that assumes a Rayleigh distribution of wave
heights, a rogue wave should be an extremely rare event \cite{r1}. 
The measurement of an ocean rogue wave (the Draupner wave) in 1995
initiated an intense interest in the subject of extreme events. It 
has been found that ocean rogue waves occur more regularly than the
statistical description predicts \cite{r1}, and a number of alternative
mechanisms for the formation of rogue waves has been produced \cite{r1}.
One such approach is through the derivation of simple modulation equations 
such as the nonlinear Schr\"odinger  (NLS) equation from the underlying 
equations of motion \cite{NLS}. The Peregrine soliton solution of the 
focusing NLS equation sits atop a finite background, and is localized in 
both space and time \cite{Peregrine}. The maximum amplitude of the 
Peregrine soliton is three times greater than the background upon which 
it sits, and is therefore a prominent rogue wave candidate.  
Such structures have been studied in various media, including nonlinear optics
\cite{o1,o2,o3,o4}, mode-locked lasers \cite{laser}, superfluid helium 
\cite{super}, hydrodynamics \cite{hydro1,hydro2,hydro3}, Faraday surface 
ripples \cite{Faraday}, parametrically-driven capillary waves \cite{cap}, 
plasmas \cite{plasmas}, ultra-cold gases \cite{stathis} and
electrical transmission lines~\cite{lefthand}.
A unifying theme of these varied physical 
settings of rogue waves is the relevance of the NLS setting as an approximate
model equation. Rogue waves in discrete systems are far less studied. One 
example of such a study concerns rogue waves in the  integrable Ablowitz-Ladik 
lattice \cite{RogueAL}, which is known to have an exact solution that has similar
properties as the NLS Peregrine soliton. 
At the level of granular systems, the pioneering work of~\cite{sen2}
was the first one to recognize the potential of such systems for
unusually large (rogue) fluctuations in late time dynamics, in the
absence of dissipation.

The present study concerns  a different discrete system, namely phononic lattices, 
which are systems that manipulate pressure waves (as opposed to photonic latices in 
which light waves are manipulated). Arguably, one of the most prototypical phononic
lattice is the Fermi-Pasta-Ulam-Tsingou (FPUT) lattice, which describes a one-dimensional
system of masses coupled through weakly nonlinear springs \cite{FPU55}. While the 
amount of research efforts in the direction of the FPUT lattice is immense (see the book \cite{FPUbook}, 
but also the recent review \cite{PGlattice}), rogue waves in FPUT lattices have not 
been reported on, to the best of our knowledge. In the small amplitude limit, the NLS
equation is once again a valid modulation equation, suggesting that Peregrine-soliton-%
type dynamics are possible in phononic lattices.

To demonstrate that a phononic rogue wave could in principle be observed experimentally, 
we conduct a study in the case of an one-dimensional chain of beads interacting through 
Hertzian contacts, i.e. granular crystals. Over the last two decades,  granular crystals
have received considerable attention, as is now summarized in a wide range of reviews~%
\cite{nesterenko1,sen08,theocharis_review,vakakis_review, gc_review,yuli_book,ptpaper}.
Granular crystals are remarkably tunable, which permits one to access weakly 
or strongly 
nonlinear dynamic responses. At the same time, it is possible to easily access
and arrange the media in a wide range of configurations (homogeneous, periodic, chains
with impurities, chains with local resonators, disordered chains, and many others). These aspects 
make the study of granular crystals fascinating from both fundamental and applied 
perspectives~\cite{ptpaper}. 

The remainder of the paper is structured as follows. In Sec.~\ref{sec:FPU}, we 
examine a homogeneous FPUT lattice. We derive a focusing NLS equation, which 
describes the modulation of small amplitude and rapidly oscillating plane waves
in time and space. The Peregrine soliton of the NLS equation is used to initialize
the FPUT system which leads to rogue-like wave dynamics. The prediction based on 
the NLS approximation coincides with the numerical simulations of the FPUT lattice
up until the formation of the large amplitude wave. While the NLS approximation sees 
a decreasing and ``vanishing" of the large amplitude wave back towards the
background, the presence of a modulational 
instability causes the formation of outward propagating waves from the center of the 
lattice. We also explore generalized pulse like initial data (in the form of Gaussian
wave packets), which can lead to wide variety of behavior including soliton dynamics, 
breathing dynamics, and rogue wave dynamics. The amplitude of the initial
conditions ``selects" 
the type of observed dynamics. In Sec.~\ref{sec:gc} we consider a diatomic granular crystal
lattice. Using a focusing NLS equation derived as an envelope approximation
of this diatomic chain, we once again use the Peregrine soliton 
solution of the NLS equation to initialize the lattice dynamics. We find qualitatively 
similar behavior to that
reported for the homogeneous FPUT lattice for all mass ratios tested. 
A noteworthy finding is that the sensitivity to boundary effects
appears to depend on the chosen 
mass ratio. 
Sec.~\ref{sec:theend} draws conclusions and discusses future directions.


\section{Homogeneous Fermi-Pasta-Ulam-Tsingou Lattices} \label{sec:FPU}
\begin{figure}
     \centerline{\includegraphics[width =  \linewidth]{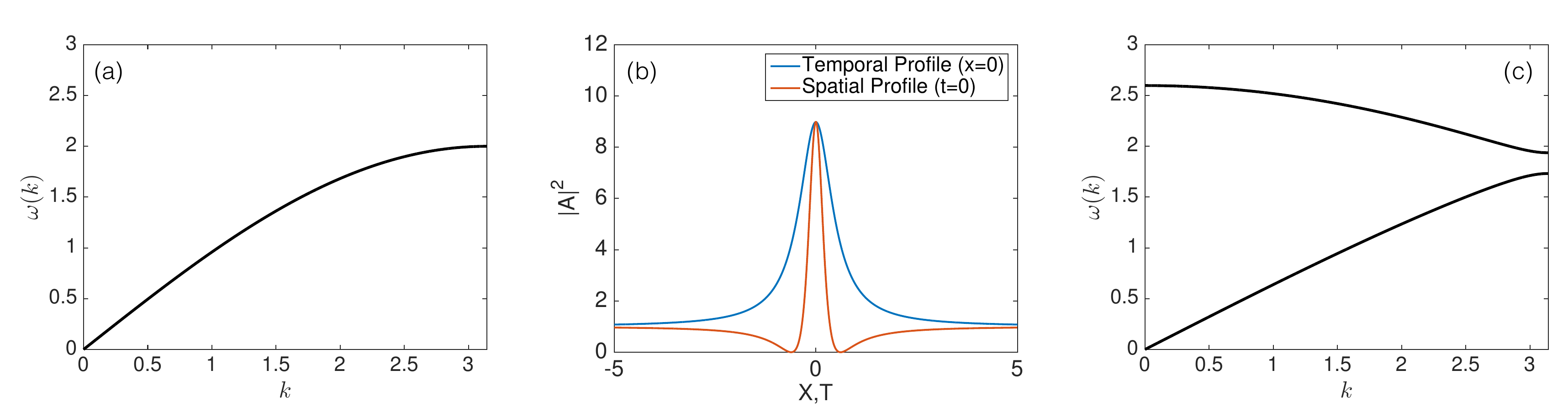}}
\caption{(a) Dispersion relation for the monomer FPUT lattice with $K_2 = 1$, 
which consists of an acoustic branch only.  (b) The spatial profile of the 
Peregrine soliton at $t=0$ and the temporal profile at $x=0$. 
(c) Dispersion relation for the dimer FPUT lattice with $K_2 = 3/2$ and 
$\rho = 0.8$, which consists of an acoustic branch (lower branch) and 
optical branch (upper branch).}
\label{disp_env}
\end{figure}

\subsection{Theoretical Set-up}

The  prototypical  Fermi-Pasta-Ulam-Tsingou (FPUT) lattice has the form
\begin{equation} \label{FPU}
 \ddot{u}_n =   V'( u_{n+1} - u_{n} ) -  V'( u_n - u_{n-1})
\end{equation}
with
\begin{equation}
V'(x) =  K_2 x + K_3 x^2 + K_4 x^3,
\label{chain}
\end{equation}
where $n \in I$, with $I$ a countable index set, 
and $u_n=u_n(t)\in\R$ is the displacement of the $n$th particle from equilibrium position at time $t$.
Equation~\eqref{FPU} with $I=\Z$ has the Hamiltonian
\begin{equation}
H = \sum_{n\in\Z} \frac{1}{2}  \dot{u}_n^2 + V(u_{n+1} - u_n). \nonumber
\end{equation}
%
%
The linear problem (i.e. when $K_3 = K_4=0$) is solved by
$$u_n(t) = e^{i(k n + \omega t)}$$ 
for all $k\in [0,\pi]$, where $\omega$ and $k$ are related through
the dispersion relation,
$$\omega(k)^2 = 4 K_2 \sin^{2}(k/2), $$
such that the cutoff point of the acoustic band is $2 \sqrt{K_2}$, 
see Fig.~\ref{disp_env}(a). Motivated by prior works on rogue waves
where the Peregrine soliton is used to describe the formation of 
such structures, we first derive the NLS equation from Eq.~\eqref{FPU}. 
When deriving the NLS equation as a modulation equation, one uses 
the multiple scale ansatz
\begin{equation} \label{ansatz}
u_n(t) \approx \psi_n(t) :=    \epsilon \left(B(X,T) + \left[ A(X,T) E+ c.c. \right] \right)\,,  %
\qquad E =   e^{i( k_0 n + \omega_0 t)},  \quad X=\eps( n + c t), \quad T = \eps^2 t,
\end{equation}
where $\eps \ll 1$ is a small parameter, effectively parametrizing the
solution amplitude (and also its inverse width). Directly substituting  
this ansatz into Eq.~\eqref{FPU} and equating the various orders of 
$\epsilon$ leads to the dispersion relation $\omega_0 = \omega(k_0)$,
at $\OO(\veps)$
the group velocity relation $c = \omega'(k_0)$, at $\OO(\veps^2)$ 
and the nonlinear Schr\"odinger equation
\begin{equation}\label{NLS}
 i \pa_T A(X,T) + \nu_2 \pa_X^2 A(X,T) + \nu_3 A(X,T)|A(X,T)|^2 = 0,
\end{equation}
at $\OO(\veps^3)$, 
where $\nu_2 = -\omega''(k_0)/2 > 0$ and $\nu_3$ is a lengthy wavenumber-dependent 
expression. Full details of the derivation of the NLS equation starting 
from Eq.~\eqref{FPU}, including the higher-order terms of the ansatz,
can be found e.g. in \cite{Schn10,Huang1,Huang2}. Since we seek standing wave 
solutions, we choose the wavenumber to be at the edge of the acoustic band
$k_0 = \pi$, such that the group velocity vanishes, $\omega_0 = 2 \sqrt{K_2}$, 
and
$$\left.\nu_3\right|_{k_0 = \pi} =  \frac{4}{K_2 \sqrt{K_2}}(3 K_2 K_4 - 4K_3^2) = b.$$
Since $\nu_2 > 0$, the NLS equation~\eqref{NLS} will be focusing if $b>0$. 
For our numerical computations, we consider the case example of $K_2 = K_4=1$ 
and $K_3 = 1/\sqrt{2}$ such that  $\nu_2 = 1/4$ and $b=4$. The equation for
$B(X,T)$ is defined in terms of $A(X,T)$,
 \begin{equation}  \label{beq}
 \pa_X B(X,T)= \frac{ 4 K_3 (1-\cos(k))}{ (\omega'(k))^2  -  (\omega'(0) )^2 }  | A(X,T) |^2.
 \end{equation}

\subsection{Peregrine Initial Data}

The focusing NLS equation has the one-parameter family of Peregrine soliton 
solutions \cite{Peregrine} given by
\begin{equation} 
\label{pere}
A(X,T) =  \sqrt{ \frac {P_0}{\nu_3}  } \left(1 - \frac{   4( 1 + 2 P_0 \, i T ) }%
{1 + \frac{2}{\nu_2} P_0 X^2 + 4 P_0^2 T^2 } \right) e^{i P_0 T},
\end{equation}
where $P_0>0$ is an arbitrary parameter. This solution is localized in space  
and time and has a maximum (located at $(x,t)=(0,0)$) that is three times greater
than its background, which are the features we desire to describe a rogue wave,
see Fig.~\ref{disp_env}(b). Using the Peregrine soliton for the envelope function
and a wave number $k_0 = \pi$, $K_2 = K_4=1$ and $K_3 = 1/\sqrt{2}$ we arrive at 
the following approximation
\begin{eqnarray}\label{approx_rogue}
u_n(t)  &=& 
\frac{\sqrt{\eta} }{2}  \left(1 - \frac{   4( 1 + 2 i \eta \,  t ) }%
{1 + 8 \eta n^2 + 4 (\eta t)^2 } \right) e^{i( \pi n +   ( 2 + \eta )t )   }  +  c.c. %
+ \epsilon B(\epsilon n, \epsilon^2 t) \, , \qquad \eta = P_0 \epsilon^2,  
\end{eqnarray}
where $B$ is defined in Eq.~\eqref{beq}. It will be convenient to represent 
the solution in the strain variable formulation, that is, $y_n = u_{n+1}-u_{n}$
since the term $B$ in the ansatz, which introduces a linear slope, will vanish. 
The parameter $\eta = P_0 \epsilon^2 >0 $ selects the background amplitude 
(since  $|y_n(0)| \rightarrow 2 \sqrt{\eta}$ as $n \rightarrow \pm \infty$) 
and the frequency of oscillation $2+ \eta$, which lies above the cutoff of 
the acoustic band $\omega_0=2$.

%
%

\begin{figure}
\centerline{\includegraphics[width = .8 \linewidth]{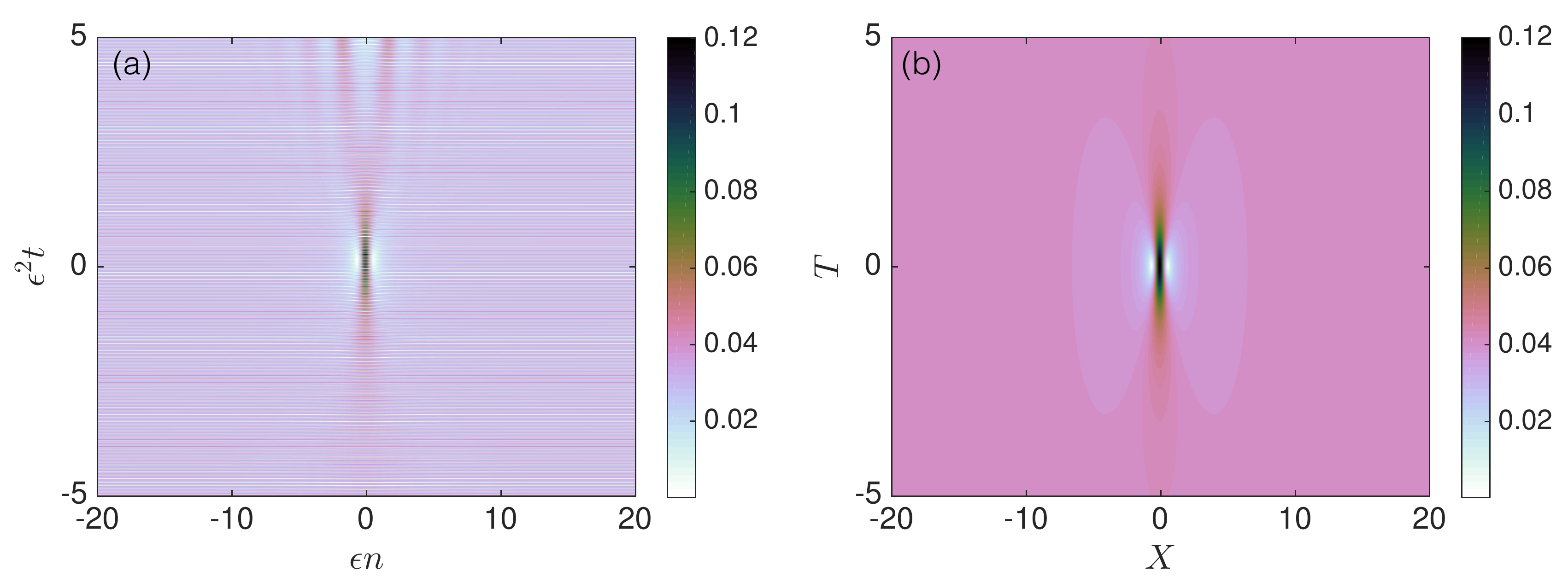} }
\caption{(a) Simulation of Eq.~\eqref{FPU} with $\epsilon=0.02$, $X \in [-40, 40]$ and  $T \in [-5,5]$ that is initialized with Eq.~\eqref{ansatz} with $A$ given by Eq.~\eqref{pere} with $P_0 = 1$.
  Color intensity corresponds to the strain $|y_n(t)|$.
  Notice that the space-time evolution here and in the figures
  that follow is given in terms of the rescaled variables
  $\epsilon n$ and $\epsilon^2 t$ for space and time, respectively.
(b) Corresponding NLS prediction $2 \epsilon |A(X,T)|$. 
Note that the background amplitude of $|y_n(t_0)|$ is the 
same as $2 \epsilon |A(X,T_0)|$. 
}
\label{eps02}
\end{figure}

To test the validity of the multiscale analysis, we perform numerical simulations
of the FPUT model Eq.~\eqref{FPU} using Eq.~\eqref{approx_rogue} as initial data.  
For instance, see Fig.~\ref{eps02} for a simulation with $\epsilon=0.02$, $X \in [-40, 40]$
and $T \in [-5,5]$.  In this simulation, our initial time is $t = -5\epsilon^2$, 
such that $t=0$ should correspond to a peak at the middle node $n=0$. The simulations
are sensitive to the boundary conditions since the background is non-zero (we employ
boundary conditions that are periodic in the strain). Therefore, we take a larger 
spatial domain to reduce the influence of the boundary, since we are mainly concerned
with the core of the solution. For times before the rogue wave appears
(i.e. $t<0$) 
the FPUT dynamics is predicted by the NLS dynamics (compare Fig.~\ref{eps02}(a) 
and ~\ref{eps02}(b)). After the formation of the rogue wave, i.e., for $t>0$, the 
FPUT dynamics departs from the NLS prediction. In the FPUT case, the large amplitude 
portion of the wave breaks into smaller, but still large relative to the background, 
waves. We believe that the emergence of these waves stemming from
the Peregrine soliton core is a byproduct of the modulational instability
of the NLS background as transcribed into the FPUT lattice and as seeded
by the large amplitude perturbation induced by the wave structure.



\subsection{Gaussian Initial Data}

\begin{figure}
\centerline{\includegraphics[width =  \linewidth]{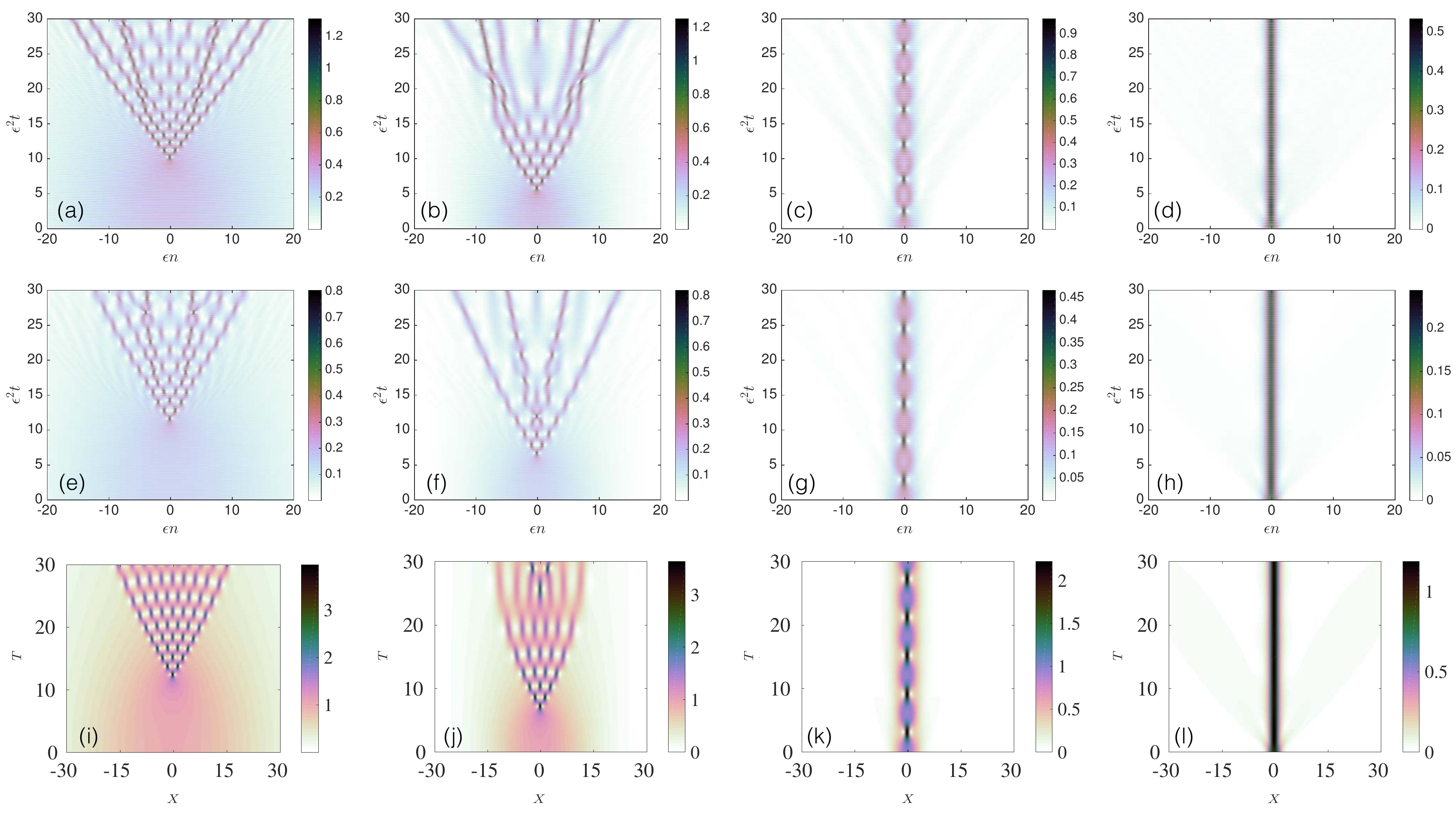} }
\caption{
Simulation of Eq.~\eqref{FPU} that is initialized with Eq.~\eqref{ansatz} 
with $A$ given by Eq.~\eqref{Gauss} with the following width parameters
$\sigma = 20.1$ for panels (a,e,i), $\sigma = 10.5$ for panels (b,f,j),
$\sigma = 2.5$ for panels (c,g,k) and $\sigma = 1.3$ for panels (d,h,l).
In panels (a)-(d) the perturbation parameter value is $\epsilon=0.1$.  
 In panels (e)-(h) the perturbation parameter value is $\epsilon=0.05$. Panels (i)-(l) 
correspond to the respective numerical simulations of the NLS equation with Gaussian initial data,
see e.g. Ref.~\cite{stathis}.
}
\label{fig:Gauss}
\end{figure}

It has  been shown through the rigorous work of~\cite{bertola}
that Peregrine-like structures are a generic by-product of the
so-called gradient catastrophe phenomenon that the (focusing) NLS is subject
to for localized initial data in the semi-classical limit.
This feature has led also to very clean recent observations of
Peregrine solitons in optical systems~\cite{suret}.
Also, at a numerical level, systematic explorations of
Gaussian initial data have led to 
rogue-like waves in the focusing NLS equation for sufficiently 
broad Gaussians \cite{stathis}. When sufficiently broad (so as
to be rescalable to the semi-classical regime), the
waves evolving through the equations of motion focus their
energy to the center in a Peregrine structure.
Even more remarkably, such initial data subsequently lead to the formation
  of an array of essentially identical (up to small corrections)
  Peregrine-like structures, arising at the poles of
  the so-called tritronqu{\'e}e solution of the Pain{\'e}v{\'e} I
  equation.
  On the other hand,
  if the Gaussian is sufficiently narrow, then a solution more akin to a
soliton forms; see the top panel of Fig.~\ref{fig:Gauss} for a few
examples. Here, we  investigate if a similar phenomenology is possible
in the FPUT lattice. More specifically, we consider initial data for
the envelope function $A(X,T)$ of the form
\begin{equation} \label{Gauss}
A(X,T=0):= A_G(X) = \exp \left(  \frac{-X^2}{ 4\sigma^{2}}  \right).
\end{equation}
%
%
In Fig.~\ref{fig:Gauss} results for simulations for the parameter values
$\sigma \in \{ 20.1, 10.5, 2.5, 1.3 \}$ for $\epsilon = 0.1$ and $\epsilon = 0.05$
are shown. Note the strong resemblance to the NLS prediction, however, 
after the main peak forms, there is noticeable distortion between the 
NLS prediction and the actual FPUT dynamics, just as the case in the 
Peregrine example in the previous subsection. In this simulation
the tails are decaying to zero, and thus any potential
boundary effects should be
minimal. These findings confirm once again the genericity of
the gradient catastrophe scenario of~\cite{bertola}, although presumably
the non-integrability of the present lattice distorts the ``Christmas-tree''
pattern of the subsequent Peregrines in comparison to the NLS paradigm.
Nevertheless, the pattern is still clearly discernible and progressively
reverts to breathing and ultimately to solitonic solutions as $\sigma$
decreases (i.e., along the horizontal direction). On the other hand,
the trend of decreasing $\epsilon$ (along the vertical
direction) makes the patterns
appear more and more ``NLS-like'' as is expected by the increased
accuracy of the NLS approximation in the limit of small $\epsilon$.

\section{Diatomic Granular Crystal} \label{sec:gc}



\subsection{Theoretical Set-up}

We now turn our attention to another variant of the FPUT model that 
considers a so-called Hertzian contact~\cite{nesterenko1,sen08} for the nonlinearity rather than the polynomial
nonlinearity considered in Eq.~\eqref{FPU}. Such a nonlinearity is 
relevant in the description of granular crystals when only considering
forces due to elastic compression between the particles. In this case, the 
model equations are
\begin{equation}
	\ddot{u}_n = \frac{A_n}{m_n}[\delta_{0,n}+u_{n-1}-u_n]_+^{p}-
	\frac{A_{n+1}}{m_n}[\delta_{0,n+1}+u_n-u_{n+1}]_+^{p}\,,
\label{eq:model}
\end{equation}
where $u_n$ is the displacement of the $n$th particle measured from its 
equilibrium position in an initially compressed chain, $m_n$ is the mass
of the $n$th particle, and $\delta_n = ({F_0}/{A_n})^{1/p}$ is a static 
displacement for each particle that arises from the static load $F_0 =
\text{const}$. For spherical particles, the exponent is $p=3/2$ and the 
parameter $A_n$, which reflects the material and the geometry of the chain's
particles, has the form
\begin{equation} \label{material}
	A_{n} = \frac{4 E_{n}E_{n+1}\sqrt{\frac{R_{n}R_{n+1}}{\left(R_{n}+R_{n+1}\right)}}}
{3E_{n+1}\left(1-\nu_{n}^2\right) + 3 E_{n}\left(1-\nu_{n+1}^2\right)},
\end{equation}
where $E_{n}$ is the elastic (Young) modulus of the $n$th particle, 
$\nu_{n}$ is its Poisson ratio, and $R_n$ is its radius. Important 
special cases of Eq.~\eqref{eq:model} include monoatomic (i.e., ``monomer'')
chains (in which all particles are identical, so $A_n=A$, $m_n = m$, and
$\delta_{0,n} = \delta_{0}$),  period-2 diatomic chains, and chains with 
impurities (e.g., a ``host'' monomer chain with a small number of ``defect''
particles of a different type).  In a monomer chain with strong precompression,
Eq.~\eqref{eq:model} is well approximated by the FPUT model Eq.~\eqref{FPU}.
To see this, let $V'(x) = -F(-x)$, and Taylor expand the nonlinearity 
$F(x) = [\delta_0 + x]^{3/2}_{+}$ about $x=0$. This leads to
\begin{equation} \label{GCcoeff}
 V'(x)  = -\delta_0^{3/2} + K_2 x + K_3x^2 + K_4x^3, \qquad  K_2 = \frac{3}{2}\delta_0^{1/2}, \quad K_3 = -\frac{3}{8} 
\delta_0^{-1/2},\quad K_4 = -\frac{3}{48}\delta_0^{-3/2}.
\end{equation}
Thus, in the small amplitude limit (where the above Taylor expansion is
valid), one can apply the same multiple scale analysis as in Sec.~\ref{sec:FPU}. 
However, in the case of the coefficients given in~\eqref{GCcoeff} the linear 
and nonlinear coefficients of the NLS equation are respectively $\nu_2 = - \omega''(\pi)/2 > 0$  
and  $\nu_3 = (K_2 K_4 - 4K_3^2)/(K_2\sqrt{K_2}) < 0$, and thus, the 
relevant NLS equation for the monomer granular crystal is the defocusing NLS. 
While this case is interesting in its own right (allowing the existence of 
NLS dark solitons, and hence dark breathers of the granular crystal \cite{dark,dark2}), 
there are no Peregrine solitons. To obtain a focusing NLS equation (and hence
the possibility of Peregrine solitons) we modify our lattice model such that
there are additional branches in the dispersion. In particular, we seek a dispersion
relation such that one branch has the opposite concavity of the acoustic branch
of the monomer chain at $k=\pi$, namely that $-\omega''(\pi)/2 < 0$. Such is the
case for the dimer granular crystal (see Fig.~\ref{disp_env}(c)) which consists 
of alternating particles of two types, so $A_n=A$, $\delta_{0,n} = \delta_{0}$, 
and the mass is $m_n = m$ for even $n$ and $m_n = M$ for odd $n$. In such a chain, 
the mass ratio $\rho =  m/M$ is the only additional parameter beyond the monomer 
case. If we introduce new variables $v_n$ to represent the displacement of the 
even particles (e.g. $v_n = u_{2k}$)  and $w_n$ to represent the displacement of
the odd particles (e.g. $w_n = u_{2k+1}$), then we can re-write Eq.~\eqref{eq:model} 
as
\begin{eqnarray}
\rho       \ddot{v}_n &=&  [1+w_{n-1} - v_n]_+^{3/2} - [1 + v_n - w_n]_+^{3/2},  \label{eq:dimer1} \\
\ddot{w}_n &=& [1+v_{n} - w_n]_+^{3/2} - [1 + w_n - v_{n+1}]_+^{3/2},   \label{eq:dimer2}
\end{eqnarray}
where we have re-scaled time and amplitude. We assume that $M>m$, such that 
the mass ratio $\rho<1$. Note that these equations under the assumption of 
small strain,
$$ |w_{n-1} - v_n| \ll 1, \qquad  |w_{n} - v_n| \ll 1 $$
reduce to the dimer FPUT lattice,
\begin{eqnarray}
\rho\ddot{v}_n &=&  V'(w_n - v_n)   - V'(v_n - w_{n-1} ),    \label{eq:dimerF1} \\
\ddot{w}_n &=& V'(v_{n+1} - w_n ) - V'(w_n - v_{n} ),    \label{eq:dimerF2}
\end{eqnarray}
where
\begin{equation*} 
 V'(x)  = K_2 x + K_3x^2 + K_4x^3, \qquad  K_2 = \frac{3}{2}, \quad K_3 = -\frac{3}{8} ,\quad K_4 = -\frac{3}{48}.
\end{equation*}
The linearized Eqs.~\eqref{eq:dimerF1} and \eqref{eq:dimerF2} (i.e. where $K_3=K_4=0$) have
solutions of the form $(v_n,w_n)^T = ( v^0, w^0)^T e^{i(kn + \omega t)} $, where $k$ and 
$\omega$ are related through the dispersion relation
\begin{equation}
\omega(k)_{\pm}^2 = K_2 \left(1 + \frac{1}{\rho} \right) \pm %
\sqrt{   \left(  1 + \frac{1}{\rho}  \right)^2 - \frac{4}{\rho} \sin^2\left(\frac{k}{2}    \right)   },
\end{equation}
where the minus and plus signs correspond to the acoustic and optical bands, 
respectively, of the dispersion relation, see Fig.~\ref{disp_env}(c). At the
wavenumber $k=\pi$ the upper cutoff frequency of the acoustic band is $\omega_{-}(\pi) = \sqrt{2 K_2}$
and the lower cutoff frequency of the optical band is $\omega_{+}(\pi) = \sqrt{2 K_2/\rho}$. Since $\rho <1$ there
is a band gap of size $\omega_{+}(\pi) - \omega_{-}(\pi) = \sqrt{\frac{2 K_2}{\rho}}( 1 - \sqrt{\rho} )  $. 
In order to find a rogue wave, we proceed in the same way as in the previous section. 
Namely, we derive a focusing NLS equation from Eqs.~\eqref{eq:dimerF1} and~\eqref{eq:dimerF2}
in order to obtain an approximation that has the Peregrine soliton as the envelope
function. This approximation should describe a rogue wave of the dimer granular 
crystal for small amplitudes. We will make use of numerical simulations to test
the role of the nonlinearity stemming from the Hertzian contact. To derive the 
NLS equation, we use the following ansatz \cite{Huang2},
\begin{eqnarray}
v_n(t) &=& \epsilon \left(B(X,T) + \left[ A(X,T) E(n,t;0,\omega_{+}(\pi)) + c.c. \right] \right), \label{ansatz2} \\
w_n(t) &=& \epsilon B(X,T), \label{ansatz2b}
\end{eqnarray}
where
$$\quad E(n,t;k_0,\omega_0) =   e^{i( k_0 n + \omega_0 t)}, \quad X = \epsilon n , \quad T=\epsilon^2 t.$$
Here, we have already selected
the plane wave at the bottom of the optical band to be modulated by the envelope function $A$, since
the notation is less cumbersome than in the general wavenumber case. Substitution of this ansatz
into Eqs.~\eqref{eq:dimerF1} and~\eqref{eq:dimerF2} leads to the focusing NLS equation at order $\epsilon^3$
\begin{equation}\label{NLS2}
i \pa_T A(X,T) + \nu_2 \pa_X^2 A(X,T) + \nu_3 A(X,T)|A(X,T)|^2 = 0, \qquad %
\nu_2 = -\frac{\omega_+''(\pi)}{2}, \quad  \nu_3 =\frac{K_2^2 \omega_{+}(\pi)}{2}  (  3 K_2 K_4 - 4 K_3 ).
\end{equation}
Note that, since  $\nu_3 < 0$ and $- \omega_+''(\pi)/2 <0$,
both $\nu_2$ and $\nu_3$ are negative such that Eq.~\eqref{NLS2} 
is focusing. The function $B(X,T)$ is defined in terms of $A(X,T)$
via
$$   \partial_X B(X,T) =  - \frac{4K_3}{K_2} |A(X,T)|^2.$$
Since $\nu_2$ and $\nu_3$ are negative, the Peregrine soliton for 
Eq.~\eqref{NLS2} is the same as of Eq.~\eqref{pere} but with
the appropriate sign changes:
%
\begin{equation} \label{Pere2}
A(X,T) =  \sqrt{- \frac {P_0}{\nu_3}}  \left(1 - \frac{   4( 1 - 2 P_0 \, i T ) }%
{1 - \frac{2}{\nu_2} P_0 X^2 + 4 P_0^2 T^2 } \right) e^{- i P_0 T}.
\end{equation}
Substituting this expression into Eq.~\eqref{ansatz2} leads to a 
plane wave that oscillates with temporal frequency $\omega_+(\pi) - \epsilon^2 P_0$ 
(and hence lies within the band gap of the spectrum, since $\epsilon \ll 1$) 
that is modulated by a Peregrine soliton. 

\subsection{Peregrine Initial data}

\begin{figure}
\centerline{
\includegraphics[width =  \linewidth]{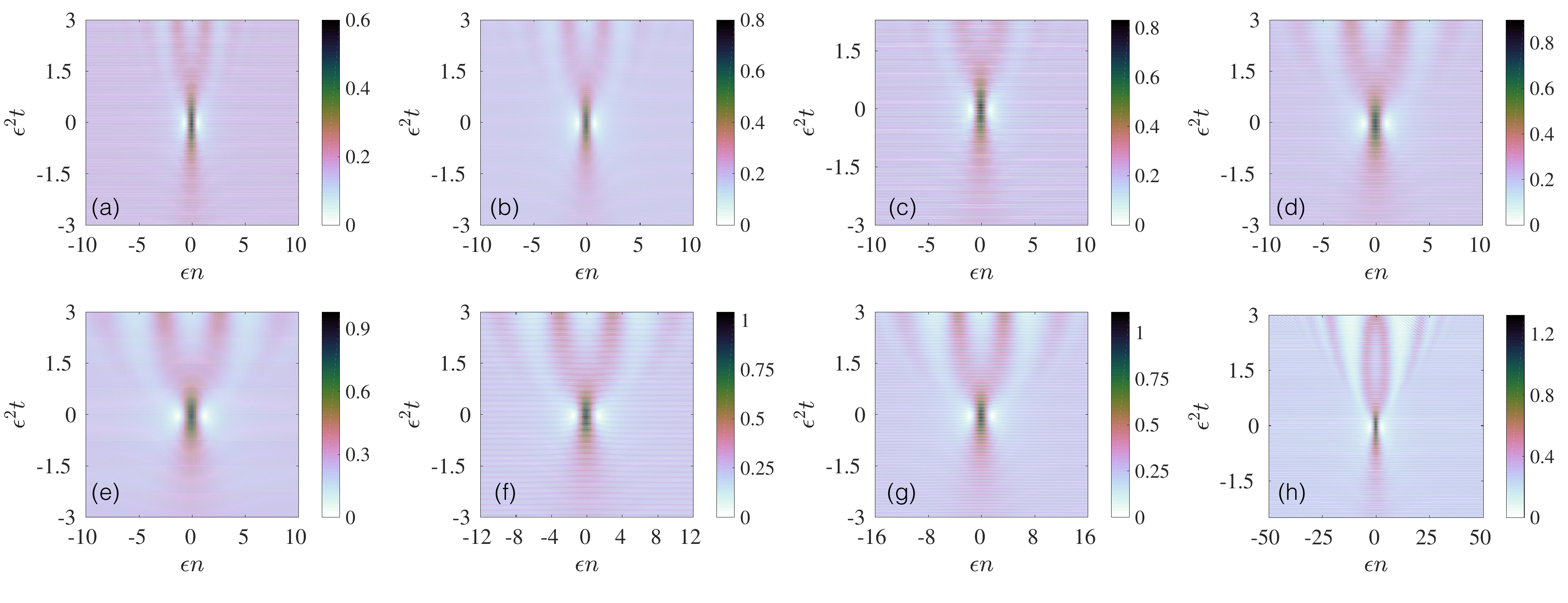} }
\caption{
Simulation of Eq.~\eqref{eq:model} that is initialized with Eq.~\eqref{ansatz2} and Eq.~\eqref{ansatz2b} with $A$
given by Eq.~\eqref{Pere2} with $P_0 = 1$,  $\epsilon=0.5$, $X \in [-40, 40]$ and  $T \in [-5,5]$. Color intensity
corresponds to the strain $|y_n|$.
The values of the mass ratio parameter are
(a) $\rho = 0.1$,
(b) $\rho = 0.2$, 
(c) $\rho = 0.3$, 
(d) $\rho = 0.4$, 
(e) $\rho = 0.5$, 
(f) $\rho = 0.6$, 
(g) $\rho = 0.7$, 
(h) $\rho = 0.9$, 
 }
\label{fig:vary_ratio}
\end{figure}

We conduct a number of simulations of the fully nonlinear dimer crystal 
model Eq.~\eqref{eq:model} using the ansatz in  Eqs.~\eqref{ansatz2}
and \eqref{ansatz2b} for various mass ratios. The results are summarized
in Fig.~\ref{fig:vary_ratio}. For small values of the mass ratio $\rho$,
the dynamics are similar to the monomer FPU chain studied above. There is
the appearance of a large amplitude peak, seemingly out of nowhere, but 
then rather than disappearing ``without a trace", as the NLS Peregrine
soliton predicts, the large amplitude portion of the wave breaks into 
smaller, but still large relative to the background, waves (compare 
Fig.~\ref{fig:vary_ratio} and Fig.~\ref{eps02}). The same feature persists 
for larger mass ratios, however, the secondary pulses become broader.
This is part of the manifestation of the modulational instability
of the corresponding background. For sufficiently large mass ratios
$\rho$, more waves seem to emerge as a result of the instability
and the time scale of their interaction appears to be shorter.

We 
have also observed a substantial sensitivity to boundary conditions and
a rapid propagation of the resulting excitations, reflecting from the boundary
towards the core of the Peregrine structure.
It is relevant to note here that 
for uncompressed granular crystals, solitary waves are found to exist at
special mass ratios (the so-called anti-resonances), and severe wave attenuation
occurs at other special mass ratios (the so-called resonances),~\cite{JKdimer,vak2,Jayaprakash1,Jayaprakash2}. It would be particularly interestsing
to explore whether such phenomena have an analogue in the case of
precompressed diatomic granular crystals and whether they have any
implications towards the formation of the Peregrine solitons.
Future studies concerning resonances and anti-resonances of precompressed diatomic
granular chains would therefore not only be interesting in their own right, but 
might also help explain the observed deviations from granular crystal dynamics 
and the NLS predictions.

%

\section{Discussion and future directions} \label{sec:theend}

In the present study, we have definitively illustrated the
potential of phononic lattices to support rogue wave structures.
Our preliminary considerations focused
on the FPUT lattice as a prototypical example where rogue waves could be
excited by using the Peregrine soliton solution of the derived NLS equation
as initial data. For sufficiently wide Gaussians, we also found rogue-wave
patterns in line with the universality of the gradient catastrophe
mechanism suggested by~\cite{bertola}.
However, for Peregrine and Gaussian initial data, the formation
of the large amplitude structures led eventually
to deviations in the FPUT dynamics from
the expected predictions of the NLS approximation. While part of the 
observed discrepancies may be attributed to boundary effects, the
predominant reason for this phenomenology is the presence of
the modulational instability for the background on top of which the
Peregrine structure is formed. We also considered
a diatomic granular crystal to demonstrate that rogue wave dynamics is possible
in a system that is highly accessible in experiments
in a space-time resolved way~\cite{gc_review}. A key challenge in that regard
concerns
the large scales considered in this paper (where the NLS approximation is valid)
leading to large lattices. However, it may be interesting to try
relevant ideas in smaller lattices; some studies have considered lattices
as large as $N=81$ nodes \cite{Boechler2009}, or
even $N=188$ nodes in~\cite{dev}.
While this paper establishes important
first steps for the realization of phononic rogue waves, future theoretical studies
should consider further steps in some of these directions;
another important one involves
the suitable initialization with Peregrine-like initial data, as these
lattices permit considerable control e.g. over driving the boundaries,
but are less amenable to a distributed initialization over the entire chain.
Such topics are presently under consideration and will be reported
in future publications.

\section*{Acknowledgements} PGK gratefully acknowledges discussions with S.~Sen 
at an early stage of this work. This material is based upon work supported by the 
National Science Foundation under Grant No. DMS-1615037.
PGK gratefully acknowledges support from the US-AFOSR
under FA9550-17-1-0114.


\end{document}